\documentclass{article}

\usepackage[nonatbib, preprint]{neurips_2024}
\usepackage[utf8]{inputenc}    
\usepackage[T1]{fontenc}       
\usepackage{hyperref}          
\usepackage{url}               
\usepackage{booktabs}          
\usepackage{amsfonts}          
\usepackage{nicefrac}          
\usepackage{microtype}         
\usepackage{xcolor}            
\usepackage{graphicx}          
\usepackage{amsmath}           
\usepackage{amssymb}           
\usepackage{multirow}          
\usepackage{algorithm}         
\usepackage{algorithmic}       
\usepackage{enumitem}          
\usepackage{listings}

\usepackage[frozencache,cachedir=.]{minted}

\setminted[json]{breaklines}

\lstdefinestyle{prompts}{
    basicstyle=\ttfamily\small, 
    breaklines=true,            
    breakatwhitespace=false,    
    postbreak={},               
    frame=none,               
    keepspaces=true,            
    showstringspaces=false,      
}

\title{GOD model: \\Privacy Preserved AI School for Personal Assistant}

\author{%
  PIN AI Team$^\star$ \\
  \href{https://www.pinai.io/}{pinai.io} \\
  San Francisco, CA \\
  \texttt{www.pinai.io} \\
}

\begin{document}

\maketitle

{\def\thefootnote{$^\star$}\footnotetext{List of contributors: Bill Sun, Gavin Guo, Regan Peng, Boliang Zhang, Shouqiao Wang, Laura Florescu, Xi Wang, Davide Crapis, and Ben Wu. Please contact Bill Sun (\texttt{bill@pinai.io}) and Gavin Guo (\texttt{gavin@pinai.io}) for further questions.}\def\thefootnote{\arabic{footnote}}}

\begin{abstract}
Personal AI assistants (e.g., Apple Intelligence, Meta AI) offer proactive recommendations that simplify everyday tasks, but their reliance on sensitive user data raises concerns about privacy and trust. To address these challenges, we introduce the \textit{Guardian of Data} (GOD), a secure, privacy-preserving framework for training and evaluating AI assistants directly on-device. Unlike traditional benchmarks, the GOD model measures how well assistants can anticipate user needs—such as suggesting gifts—while protecting user data and autonomy. Functioning like an “AI school”, it addresses the cold start problem by simulating user queries and employing a curriculum-based approach to refine the performance of each assistant. Running within a Trusted Execution Environment (TEE), it safeguards user data while applying reinforcement and imitation learning to refine AI recommendations. A token-based incentive system encourages users to share data securely, creating a data flywheel that drives continuous improvement. 
Specifically, users "mine" with their data, and the mining rate is determined by GOD’s evaluation of how well their AI assistant understands them across categories such as shopping, social interactions, productivity, trading, and Web3. By integrating privacy, personalization, and trust, the GOD model provides a scalable, responsible path for advancing personal AI assistants. For community collaboration, part of the framework is open-sourced at \url{https://github.com/PIN-AI/God-Model}.
\end{abstract}

\begin{figure}[!ht]
    \centering
    \vspace{-8mm}
    \includegraphics[width=0.95\linewidth]{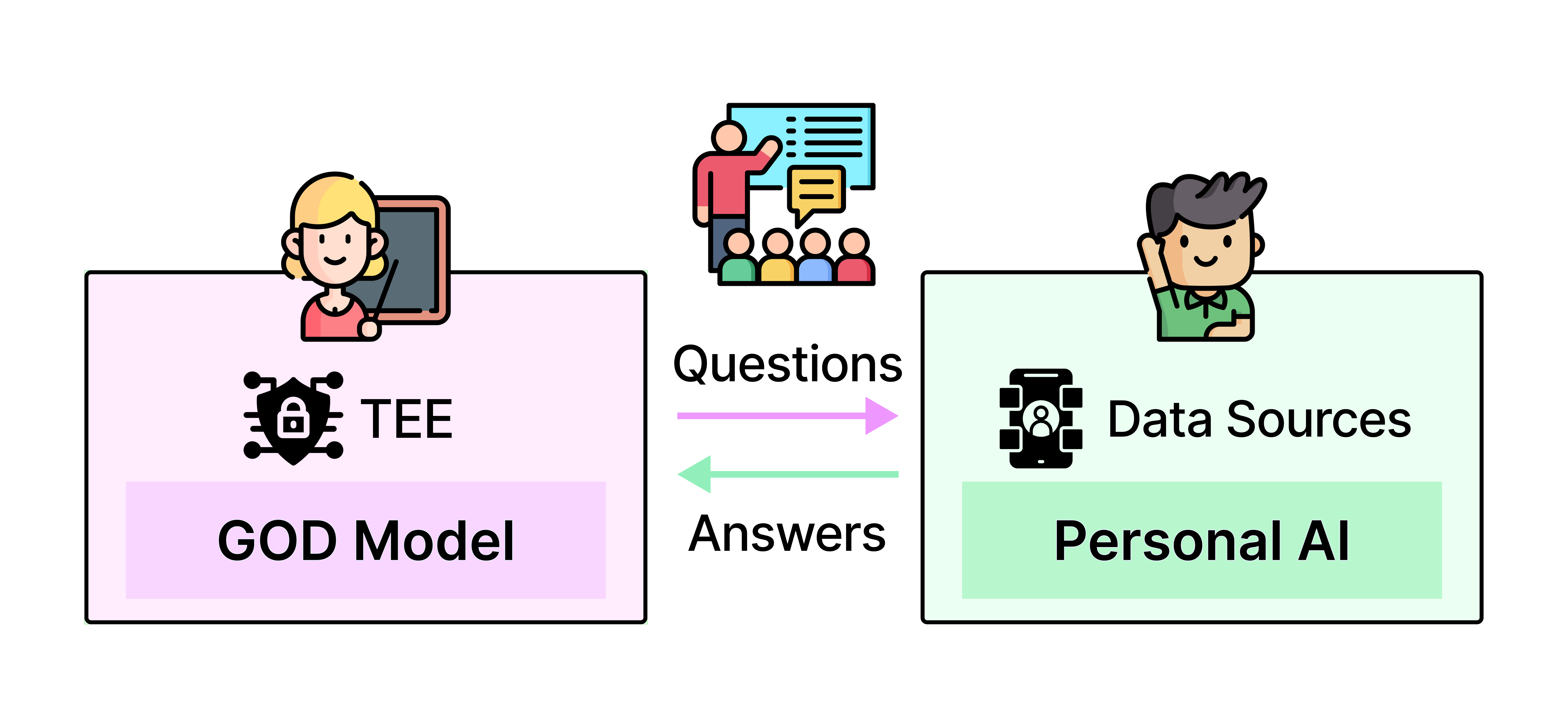}
    \vspace{-4mm}
    \caption{The GOD model, running in a secure TEE, generates realistic user queries. The Personal AI on the user’s device processes private data to answer the queries. The GOD model verifies and grades those answers through internal checks or external APIs while ensuring privacy.}
\end{figure}

\section{Introduction}
On-device AI agents show promise for contextual understanding, privacy controls, and responsive user interactions. However, as these agents move beyond \emph{responding} to explicit user queries and begin to \emph{anticipate} future needs, they face a trust barrier. Recommending a birthday gift, notifying users of an upcoming flight deal, or suggesting a vacation trip plan can all be very helpful, but these proactive features depend on gathering, interpreting, and sometimes inferring from private data. Users may not fully expect such inferences, raising concerns about transparency, autonomy, and responsible data handling.

To address these challenges, we present the \emph{Guardian of Data} (GOD) model, a secure framework designed to evaluate and improve personal AI on device. Unlike generic benchmarks that focus on general knowledge or factual accuracy, the GOD model focuses on proactive tasks that leverage sensitive data, both to measure performance and to ensure responsible usage. By operating within a Trusted Execution Environment (TEE), the framework maintains strong privacy guarantees while performing quality assessments. The GOD model generates its own `exam papers', curriculum-based queries of varying difficulty, and distributes them to personal AI agents around the world. The answers of each assistant are based on different types of user data (e.g., emails, social networks, receipts), yet the classification and analysis occur securely, with the raw user data never leaving the device. Over time, the model can act as a teacher, offering step-by-step demonstrations to help each on-device assistant refine its behavior.

A key advantage of this approach is its ability to mitigate the cold start problem in personalization and recommendation. The GOD model generates realistic user queries and usage patterns, enabling newly created personal AIs to train before engaging with real users. Additionally, a token-based incentive mechanism promotes secure data sharing, driving a data flywheel effect. This continuous cycle of data collection, evaluation, and refinement enhances personalization while preserving user autonomy and security. Below, we outline our main contributions:

\begin{enumerate}[leftmargin=2.4em]  
    \item \textbf{TEE-Based Secure Evaluation:} We introduce a trusted execution environment (TEE) that enables robust scoring and feedback while safeguarding user data from unauthorized access.  
    \item \textbf{Curriculum-Based Assessment:} Our approach evaluates AI proficiency by starting with factual queries and progressively introducing more complex, context-rich, preference-based recommendations.  
    \item \textbf{Data Value Estimation:} We propose a structured method to quantify the impact of personal data on recommendation quality, helping stakeholders balance personalization benefits with privacy concerns.  
    \item \textbf{Anti-Gaming Safeguards:} By implementing cross-verification, identity checks, and anti-fraud mechanisms, we ensure the integrity of test results and rewards, promoting fair and transparent evaluation.  
\end{enumerate}

These elements collectively establish a framework for both \emph{assessing} and \emph{enhancing} on-device AI systems, enabling them to deliver proactive support without undermining user trust.

\section{Related Work}  
\label{sec:related_work}  

The evaluation of personalized large language models (LLMs) has been a growing area of research, with existing works primarily focusing on measuring how well models adapt to user-specific contexts, styles, or preferences. However, most benchmarks emphasize \emph{reactive} personalization—assessing how well an LLM responds to explicit user prompts—rather than evaluating \emph{proactive} personalization, which is crucial for personal AI assistants.

\subsection{Existing Personalization Benchmarks}

Benchmarks such as \textbf{LaMP} \cite{salemi2024lamplargelanguagemodels} and \textbf{LongLaMP} \cite{kumar2024longlampbenchmarkpersonalizedlongform} provide user-specific ground-truth datasets, enabling direct comparisons between model outputs and actual user-authored content. Others, such as \textbf{PersoBench} \cite{afzoon2024persobenchbenchmarkingpersonalizedresponse} and \textbf{PrefEval} \cite{anonymous2024do}, test persona consistency and adherence to stated preferences. Although these datasets help quantify personalization gains, they rely on static corpora, lack real-time adaptation, and do not assess privacy-preserving or on-device AI performance.

In personalized recommendation, benchmarks like \textbf{MovieLens} \cite{harper2015movielens} and \textbf{Amazon Reviews} \cite{ni-etal-2019-justifying} evaluate user item predictions but assume centralized data collection, making them unsuitable for privacy-focused, on-device applications. More recent approaches, such as \textbf{Ranking-TAGER} \cite{anonymous2024premium}, integrate user tagging and ranking-based personalization but lack trusted execution environments (TEEs) or mechanisms to prevent fraudulent data submissions. Similarly, \textbf{PersonalLLM} \cite{zollo2024personalllmtailoringllmsindividual} investigates latent user preferences, but does not address privacy concerns or adversarial data manipulation.

\subsection{Limitations of Existing Evaluation Metrics}

Traditional evaluation metrics for personalization primarily assess response accuracy, coherence, and user preference alignment. \textbf{Intrinsic metrics} such as \textbf{BLEU} \cite{papineni-etal-2002-bleu}, \textbf{ROUGE} \cite{lin-2004-rouge}, and \textbf{METEOR} \cite{banerjee-lavie-2005-meteor} measure textual similarity but do not account for the proactive nature of personal AI assistants. More personalized evaluation methods, such as \textbf{Win Rate} \cite{zhang2024personalizationlargelanguagemodels}, \textbf{Hits@K} \cite{mazaré2018trainingmillionspersonalizeddialogue}, and \textbf{Word Mover’s Distance} \cite{ghorbani2019automaticconceptbasedexplanations}, provide better alignment with user preferences, but still focus on \emph{reactive} tasks rather than anticipatory interactions.

\textbf{Extrinsic metrics}, including \textbf{recall}, \textbf{precision}, and \textbf{normalized discounted
cumulative gain} (NDCG) \cite{wang2013theoreticalanalysisndcgtype}, are widely used in recommendation systems but assume centralized data aggregation, making them impractical for privacy-preserving, on-device personalization. Furthermore, these metrics do not quantify the actual \emph{value} of user data in personalization, a critical aspect to balance personalization with privacy.

\section{System Components}  
We outline four key components—(a) the GOD model, (b) the Personal AI, (c) Data Connectors, and (d) the HAT Node—that collectively preserve user privacy while enabling proactive AI recommendations.

\begin{figure}[!ht]
    \centering
    \includegraphics[width=\linewidth]{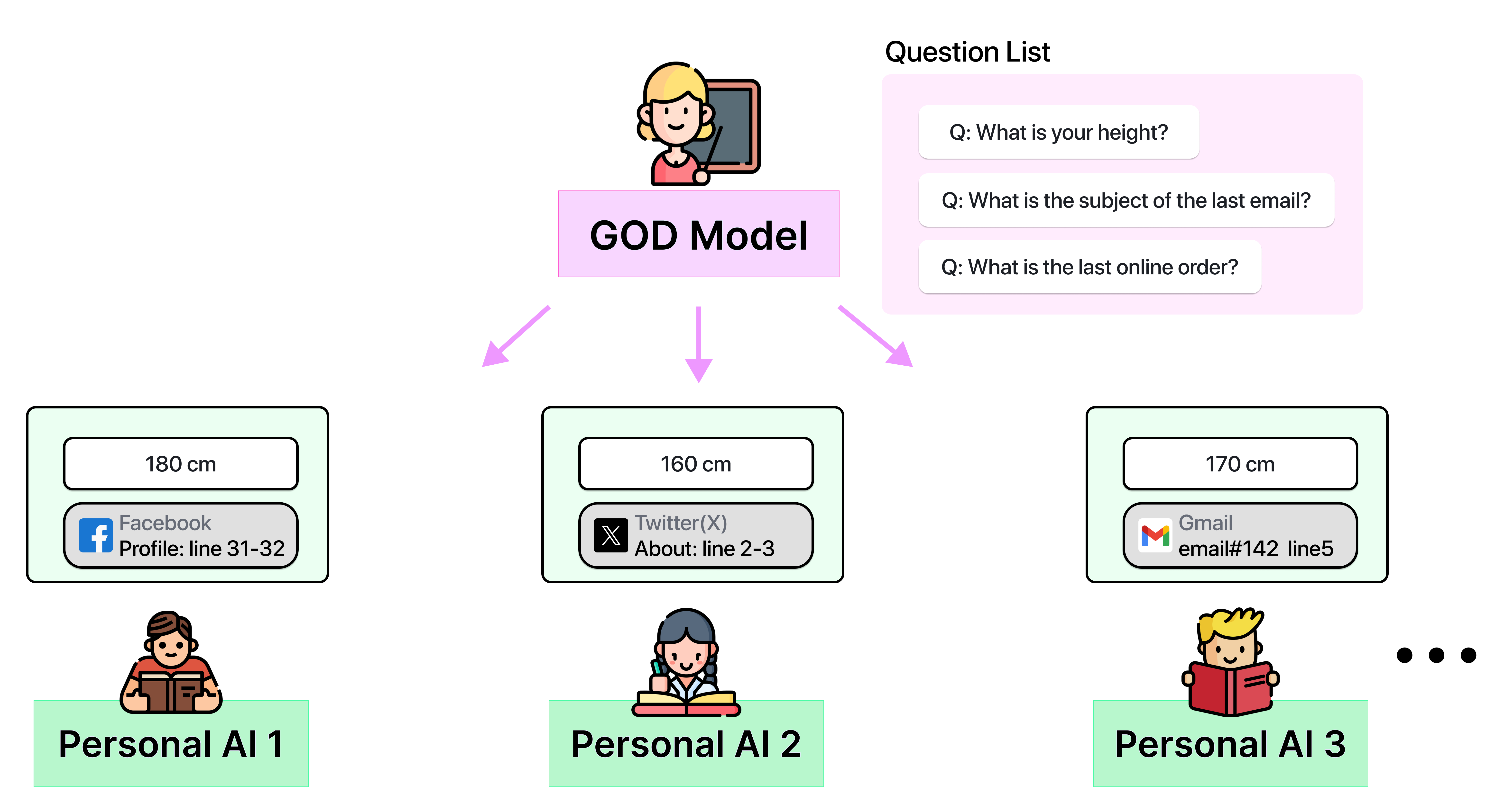}
    \caption{The GOD model evaluates the Personal AI by posing personal questions (e.g., height, last email subject, recent online orders). The Personal AI uses only private, on-device data to respond without revealing raw information.}
\end{figure}

\begin{figure}[!ht]
    \centering
    \includegraphics[width=\linewidth]{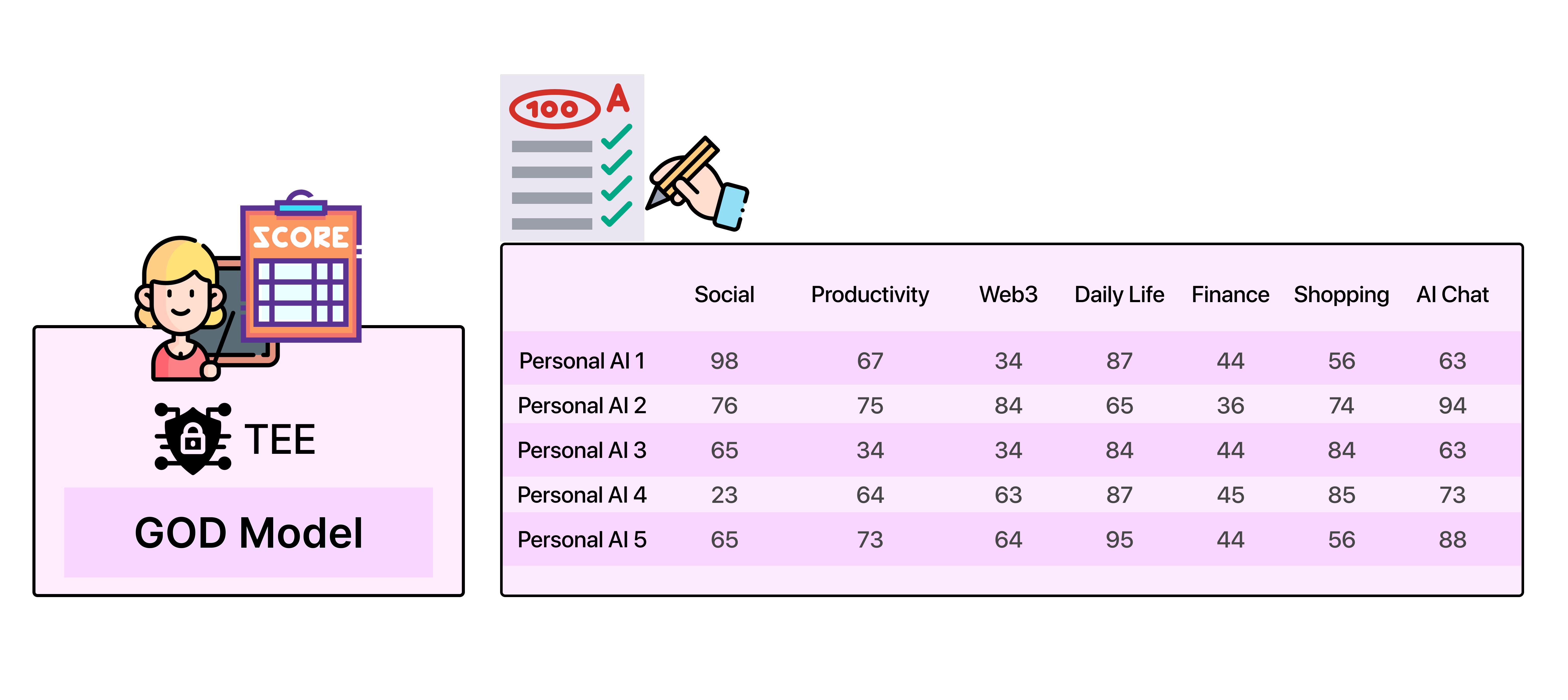}
    \vspace{-8mm}
    \caption{The GOD model uses Data Connectors to securely access user data within a TEE, then scores the Personal AI’s responses in multiple categories.}
\end{figure}

\textbf{GOD Model:}  
The GOD model runs inside a Trusted Execution Environment (TEE), simulating real-world tasks—such as booking travel or choosing gifts—to evaluate how each AI uses personal data. It verifies responses using Data Connectors that securely access user-private data in the TEE, then assigns performance scores based on correctness and proactivity. An “Exam Designer,” powered by a powerful language model or rule-based logic, generates and grades questions within the TEE.

\textbf{Personal AI:}  
Each user’s Personal AI operates locally, analyzing private data (e.g., emails, social media posts, calendars) without transmitting raw information off-device. It responds to the GOD model’s queries using on-device language models with search and retrieval augmented generation. Over time, it becomes increasingly proactive—reminding users of important dates or suggesting meeting times—while still protecting personal information.

\textbf{Data Connectors:}  
These secure interfaces validate details with external services (e.g., confirming airline bookings or cryptocurrency balances) by returning minimal signals (“verified” or “incorrect”) without revealing sensitive information. Beyond common Web2 APIs (e.g., Gmail) and Web3 explorers, Data Connectors can be extended to sources like AR/VR logs or biometric sensors, prioritizing privacy through minimal data disclosure.

\textbf{HAT (Human Assessment Test) Node:}  
To prevent fraud or “Sybil” accounts, the HAT Node verifies user identities using phone checks, ID documents, or blockchain histories. Users are categorized into basic, verified, or high-trust tiers, each with corresponding reward levels.

\section{Evaluation Methodology}
\label{sec:eval_method}

Evaluating the ability of a Personal AI to leverage private data for proactive recommendations requires a structured, transparent, and privacy-preserving approach. The GOD model accomplishes this through a curriculum-based assessment, a well-defined scoring function, and explicit methods to quantify the value of personal data.

\subsection{Curriculum-Based Assessment}
The GOD model follows a curriculum evaluation that begins with basic data retrieval tasks and advances to more complex, preference-based queries. This progression reveals how well the AI evolves as it tackles increasingly sophisticated user queries:
\begin{itemize}[leftmargin=3em]
    \item \textbf{Level 1 (Easy):} Simple factual recall (e.g., "What was the subject of the user’s last email?").
    \item \textbf{Level 2 (Medium):} Cross-referencing tasks that require integrating data from multiple sources (e.g., linking calendar events with email content).
    \item \textbf{Level 3 (Hard):} Context-rich queries that demand personalized recommendations (e.g., "Suggest a restaurant based on dietary preferences and location history.").
\end{itemize}

\subsection{Scoring Function}
\label{sec:scoring}
The GOD model employs a scoring function that combines coverage, quality, and freshness:
\[
\text{Total Score} = w_C \cdot \text{Coverage} + w_Q \cdot \text{Quality} + w_F \cdot \text{Freshness}.
\]
\begin{itemize}[leftmargin=3em]
  \item \textbf{Coverage (C):} Measures the proportion of data domains (email, calendar, etc.) the AI effectively utilizes.
  \[
    C = \frac{\text{Number of Utilized Data Domains}}{\text{Total Number of Available Data Domains}}
  \]
  \item \textbf{Quality (Q):} A weighted combination of \textit{accuracy}, \textit{consistency}, and \textit{proactive success}.
  \[
    Q = \alpha \cdot \text{Accuracy} + \beta \cdot \text{Consistency} + \gamma \cdot \text{Proactive Success Rate}
  \]
  \item \textbf{Freshness (F):} A time-decay function indicating how up-to-date the AI’s data index is:
  \[
    F = e^{-\lambda t},
  \]
  \item \textbf{Weights ($w_C$, $w_Q$, $w_F$):} Define the relative influence of each component (e.g., $w_C = 0.3$, $w_Q = 0.5$, $w_F = 0.2$).
\end{itemize}
All components are normalized within the range [0,1] to ensure comparability and clarity in performance evaluation.

\subsection{Measuring the Value of Personal Data}
A key objective of our framework is to quantify the tangible impact of personal data on AI recommendations. By comparing AI performance under personalized and baseline conditions, the GOD model highlights the value of user-specific data contributes, helping developers and users balance personalization with privacy.

\begin{enumerate}[leftmargin=3em]
    \item \textbf{Task Selection:} Define common AI use cases (e.g., trip planning, email summarization).
    \item \textbf{Dual Execution:} Run AI tests in two configurations:
    \begin{itemize}
        \item \textit{Personalized Access:} AI utilizes user-specific data (purchase history, emails, preferences).
        \item \textit{Baseline (Control):} AI operates without personal data, relying only on general knowledge.
    \end{itemize}
    \item \textbf{Output Comparison:} Assess results using:
    \begin{itemize}
        \item \textit{User Feedback:} Blind ratings on relevance and satisfaction.
        \item \textit{Automated Metrics:} Evaluation using measures like ROUGE and efficiency scores.
        \item \textit{A/B Testing:} Present different outputs to user groups to gauge engagement.
    \end{itemize}
    \item \textbf{Value Quantification:} Compute the personalization gain by comparing outcomes across both settings.
\end{enumerate}

\subsection{Progressive Curriculum Examples}
\label{sec:progressive-curriculum}
In Figure \ref{fig1}, \ref{fig2}, and \ref{fig3}, we provide concrete examples at Levels 1, 2, and 3 to illustrate the GOD model’s curriculum-based progression, demonstrating how AI competency increases in complexity.

\begin{figure}[!h]
    \centering
    \begin{minted}
    [fontsize=\footnotesize]
    {json}
    {
    "level": 1,
    "difficulty": "Easy",
    "tasks": [
        {
            "task_name": "Single-Source Recall",
            "prompt": "What was the subject of the most recent email in the user’s inbox?",
            "skills_tested": "Simple lookup or keyword search within a single data domain (e.g., email).",
            "evaluation": "GOD model confirms correctness by matching the hashed or reference-verified email subject."
        },
        {
            "task_name": "Simple Personal Fact",
            "prompt": "What is the user’s next calendar event?",
            "skills_tested": "Direct retrieval from the user’s calendar data.",
            "evaluation": "Verified by checking minimal metadata (e.g., hashed event ID)."
        }
    ]
    }
    \end{minted}
    \caption{Level 1 (Easy): Direct Factual Queries}
    \label{fig1}
\end{figure}

\begin{figure}[!h]
    \centering
    \begin{minted}
    [fontsize=\footnotesize]
    {json}
    {
    "level": 2,
    "difficulty": "Medium",
    "tasks": [
        {
            "task_name": "Cross-Domain Linking",
            "prompt": "Suggest the best time to schedule a doctor’s appointment, considering both the user’s calendar and their travel plans found in email.",
            "skills_tested": "Merging multiple data sources (calendar events + email travel confirmations).",
            "evaluation": "GOD model checks whether the AI accounts for any travel conflicts and provides a valid time window."
        },
        {
            "task_name": "Preference Integration",
            "prompt": "Recommend a movie for tonight, based on the user’s favorite genres, recent streaming history, and free time available.",
            "skills_tested": "Inference of user preferences from watch logs or ratings, alignment with schedule from calendar or location data.",
            "evaluation": "GOD compares the AI’s recommendation against known user preferences, measuring potential acceptance or engagement."
        }
    ]
    }
    \end{minted}
    \caption{Level 2 (Medium): Cross-Referencing and Context}
    \label{fig2}
\end{figure}

\begin{figure}[!h]
    \centering
    \begin{minted}
    [fontsize=\footnotesize]
    {json}
    {
    "level": 3,
    "difficulty": "Hard",
    "tasks": [
        {
            "task_name": "Complex Task Coordination",
            "prompt": "Plan a surprise birthday outing for a friend, factoring in the user’s budget, the friend’s dietary restrictions, restaurant availability, and any relevant coupons in the user’s email.",
            "skills_tested": "Multi-step reasoning, personal preference inference, real-time availability checks, creative event planning.",
            "evaluation": "GOD model verifies consistency (venue meets dietary constraints), timeliness (booked in advance), and user satisfaction signals (simulated acceptance or rating)."
        },
        {
            "task_name": "Proactive Alerts",
            "prompt": "Without a direct user request, the AI suggests blocking time for finalizing a project due next week, referencing email deadlines and calendar tasks.",
            "skills_tested": "Predictive modeling of user needs, balancing helpfulness with non-intrusiveness.",
            "evaluation": "System rates suggestions on timeliness, relevance, and effectiveness in preventing scheduling conflicts."
        }
    ]
    }
    \end{minted}
    \caption{Level 3 (Hard): Nuanced Recommendations and Proactive Behavior}
    \label{fig3}
\end{figure}

As tasks escalate in complexity, the GOD model assesses both factual accuracy and the AI’s ability to integrate, infer, and proactively assist users.

\section{Anti-Gaming Protections}
\label{sec:anti_gaming}

The GOD model rewards users whose Personal AI effectively utilizes their data. However, without safeguards, the system could be exploited through fake records or coordinated fraud. To prevent this, the GOD model enforces a multi-layered anti-gaming strategy by verifying users through KYC, validating user data, and confirming provider transactions. This ensures fairness, privacy, and trust. Below are the key measures in place.

\subsection{High-Level Overview}

\begin{itemize}[leftmargin=2em]
    \item \textbf{User Verification (KYC)}
    \begin{itemize}[leftmargin=1.5em]
        \item A \emph{HAT Node} verifies user identities using on-chain records or real-world documents.
        \item Higher reward tiers are available to users with stronger verification, discouraging fake accounts and Sybil attacks.
    \end{itemize}
    
    \item \textbf{User Data Validation}
    \begin{itemize}[leftmargin=1.5em]
        \item \textbf{Real-Time Data:}  
        New events (e.g., ride receipts, email notifications) are instantly verified in a \emph{Trusted Execution Environment} (TEE), preventing backdated or fabricated records.
        \item \textbf{Older (Dumped) Data:}  
        Past records (e.g., archived emails, on-chain transactions) are hashed and stored on-device. These are periodically checked against real-time data to detect inconsistencies.
    \end{itemize}
    
    \item \textbf{Provider Data Verification}
    \begin{itemize}[leftmargin=1.5em]
        \item Platforms like rideshare and e-commerce providers submit \emph{privacy-preserving proofs} (e.g., hashed transaction IDs) to confirm real transactions.
        \item The TEE cross-references timestamps, amounts, and reference IDs to detect mismatches or duplicate claims.
    \end{itemize}
\end{itemize}

\subsection{Further Safeguards and Penalties}

\paragraph{Privacy-Preserving Proof in a TEE.}
All verification occurs inside a TEE, ensuring that raw user data stays on the device. Only pass/fail results or summarized trust scores are shared, preventing external access to personal data.

\paragraph{Verifiable vs.\ Non-Verifiable Data.}
Data that can be confirmed via APIs or cryptographic proofs—such as airline tickets, blockchain transactions, and platform-issued receipts—carry more weight than free-text or self-reported information. This discourages unverifiable submissions.

\paragraph{Cross-Verification and Random Data Injection.}
The system checks user data against real-time, historical, and provider records to identify inconsistencies. It also injects fake prompts (e.g., non-existent coupons) to test if a Personal AI incorrectly accepts false data. Previously verified data may be rechecked; contradictions result in stronger penalties than initial errors.

\paragraph{Collaborative Fraud Mitigation.}
Users attempting to manipulate the system through coordinated fake transactions are detected through external data sources and pattern analysis. When suspicious behavior is found, the GOD model halts new data ingestion for those accounts and triggers further investigation.

\paragraph{Penalties and Long-Term Monitoring.}
A single confirmed fraud case may result in a full score reset. Additionally, data sources with repeated fraud incidents are downgraded in the scoring algorithm. Ongoing monitoring and adaptive scoring make sustained cheating impractical.

\section{Future Evolution: A Simulated AI School with RL-Based Curriculum}
\label{sec:future_path}

Beyond static exams, the GOD model aims to build a dynamic reinforcement learning (RL) system where on-device personal AIs continuously improve their proactive support. In this phase, personal AIs progress from simple recall and basic suggestions to deeper reasoning and context awareness. They learn to understand user schedules, preferences, and habits and to use advanced tools (e.g., function calling for restaurant bookings) without sending raw data off the device. As they develop, these AIs can better predict needs such as scheduling, meal planning, or entertainment while keeping data private.

\subsection{Online Learning from Personalized Human Feedback}
Online learning from personalized human feedback helps language models adjust to individual preferences on edge devices with limited power. Recent methods tailor responses efficiently without heavy user input.

\textbf{Personalized Reinforcement Learning from Human Feedback (P-RLHF)} lets models match individual needs instead of assuming one size fits all. Using user IDs and feedback histories, P-RLHF builds unique profiles without explicit instructions. It also uses personalized Direct Preference Optimization (P-DPO) to generate responses that fit diverse preferences. This approach works well for both familiar and new users, showing strong adaptability and scalability~\cite{li2024personalizedlanguagemodelingpersonalized}.

Given a personalized preference dataset 
\[
\mathcal{Z}_p = \{(\mathbf{x}_i, \mathbf{y}_{i,1}, \mathbf{y}_{i,2}, \mathbf{u}_i)\}_{i=1}^N,
\]
where $\mathbf{u}_i \in \mathcal{U}$ is user information, $\mathbf{x}_i$ is the prompt, and $\mathbf{y}_{i,1} \succ \mathbf{y}_{i,2}$ means that user $\mathbf{u}_i$ prefers $\mathbf{y}_{i,1}$ over $\mathbf{y}_{i,2}$, P-RLHF learns:
\begin{itemize}
   \item A user model $f_{\mathbf{\Phi}}: \mathcal{U} \rightarrow \mathbb{R}^{T_u \times d}$ that maps user data to embeddings $\mathbf{e}_u$
   \item A personalized language model $\mathcal{P}_{\mathbf{\Phi}}(\cdot|\mathbf{x},\mathbf{u})$ that generates responses based on both prompt $\mathbf{x}$ and user $\mathbf{u}$
\end{itemize}

The objective for P-DPO is:
\begin{align*}
\min_{\mathbf{\Phi}} -\mathbb{E}_{(\mathbf{x},\mathbf{y}_1,\mathbf{y}_2,\mathbf{u})\sim\mathcal{Z}_p} &\Big[\alpha \log \sigma\Big(\beta \log \frac{\mathcal{P}_{\mathbf{\Phi}}(\mathbf{y}_1|\mathbf{x},\mathbf{u})}{\mathcal{P}_{\mathbf{\Phi}_0}(\mathbf{y}_1|\mathbf{x})} - \beta \log \frac{\mathcal{P}_{\mathbf{\Phi}}(\mathbf{y}_2|\mathbf{x},\mathbf{u})}{\mathcal{P}_{\mathbf{\Phi}_0}(\mathbf{y}_2|\mathbf{x})}\Big) \\
&+ (1-\alpha) \log \sigma\Big(\beta \log \frac{\mathcal{P}_{\mathbf{\Phi}}(\mathbf{y}_1|\mathbf{x},\mathbf{u}_0)}{\mathcal{P}_{\mathbf{\Phi}_0}(\mathbf{y}_1|\mathbf{x})} - \beta \log \frac{\mathcal{P}_{\mathbf{\Phi}}(\mathbf{y}_2|\mathbf{x},\mathbf{u}_0)}{\mathcal{P}_{\mathbf{\Phi}_0}(\mathbf{y}_2|\mathbf{x})}\Big)\Big]
\end{align*}

Here, $\alpha \in [0,1]$ balances specific and general preferences, $\beta > 0$ controls deviation from the pre-trained model $\mathcal{P}_{\mathbf{\Phi}_0}$, and $\mathbf{u}_0$ denotes no user data. This setup supports learning personalized and generic response patterns.

\textbf{Knowledge Graph Tuning (KGT)} provides real-time personalization using knowledge graphs without changing model parameters. KGT extracts triples of personalized knowledge from user queries and feedback to build tailored representations. By optimizing the graph instead of model weights, it reduces memory use and latency while boosting personalization. KGT has improved efficiency and scalability by up to 61\% compared to traditional methods~\cite{sun2024knowledgegraphtuningrealtime}.

\textbf{Critique and Revise (CnR)} refines the responses using natural language feedback. Rather than relying solely on ranking signals, CnR learns from both praise and criticism, improving response quality by up to 65.9\% after just a few feedback rounds. This method is ideal for data-limited environments and edge LLMs where efficiency and adaptability are crucial~\cite{jin2023dataefficientalignmentlargelanguage}.

\subsection{Imitation Learning and RL in Practice}
The framework uses a two-step approach that combines imitation learning (IL) and reinforcement learning (RL):
\begin{enumerate}
\item \textbf{Teacher Demonstrations.} A larger LLM within the TEE shows how to handle tasks that combine different data sources (e.g., calendars and dietary restrictions) with user context (e.g., budget or location). The on-device AI watches these examples and mimics them to form an initial policy.
\item \textbf{Critic and Policy Learning.} A separate critic model in the TEE provides reward signals based on user satisfaction. It rewards helpful suggestions and penalizes intrusive ones. Using methods like P-DPO, the on-device AI gradually refines its policy for long-term benefit.
\end{enumerate}

These steps help each personal AI improve over time, balancing timely and relevant suggestions with avoiding unnecessary notifications. Entropy-based exploration further encourages the AI to try new strategies instead of sticking with safe but limited ones.

\subsection{Example: Proactive Dinner Reservations}
Consider a user who typically dines out on Fridays. The GOD model first detects this pattern from the user’s data (e.g., calendar events and receipts). Next, the LLM teacher outlines a step-by-step process: checking restaurant availability, considering dietary needs, and timing the suggestion to avoid interrupting work. The on-device AI mimics this process and refines it using RL, guided by user feedback. Over time, AI becomes skilled at offering the right dining options at the right time, while keeping sensitive data on device.

Through continuous learning, the GOD model evolves from static Q\&A to an “AI school”. Personal AIs grow with user needs, maintain privacy, and learn from high-level feedback. Ultimately, they can predict, assist, and coordinate various tasks, setting a new standard for proactive, privacy-focused intelligence.

\section{Conclusion}
\label{sec:conclusion}
We have presented the Guardian of Data (GOD) framework to securely evaluate, train, and improve personal AI systems on devices aimed at proactive recommendations. By introducing curriculum-based tests, trusted execution for data safety, and anti-gaming mechanisms, the GOD model addresses the battle between personalization and privacy. It clarifies how personal data improves recommendation quality and provides a blueprint for iterative learning, both through pre-deployment simulation and post-deployment adaptation.

Looking ahead, we plan to refine preference modeling, explore more advanced reinforcement learning techniques, and incorporate metrics that better capture long-term user satisfaction. These extensions will help personal AIs seamlessly anticipate user needs, identify meaningful opportunities for assistance, and remain transparent and respectful of user autonomy. In doing so, the GOD model paves the way for truly trustworthy AI companions, ones that operate as informed partners while preserving the sanctity of personal data.

\bibliographystyle{unsrt}
\bibliography{ref}

\newpage

\appendix
\section{Appendix}
\subsection{Example Prompt for GOD model Exam Question List}

\begin{lstlisting}[style=prompts, 
caption=Prompt for instruction annotation, 
label=lst:prompt_scoring]
## **Hand-Coded Question Design for Personal AI Memory System**

### **Objective**
Each question must meet the following criteria:

1. **Easy to Grade:**  
   - The question should be graded by an **Intel SGX TEE compute engine**.  
   - Grading involves checking factual existence in the **user's data** (via API or data dump).  
   - The system searches JSON data, applies hard-coded Python summarization, and confirms findings using a **local LLM R1 1.5B model**.
   - The grading process ensures:
     - **YES** Data exists and matches source.  
     - **NO** Data is missing or unavailable, and a **TEE Grader GOD model** suggests new data sources.

2. **Useful Personalization for Transactions:**  
   - The question should predict **future transactions** for AI recommendations.  
   - Example: *"Your friend's birthday is in 7 days. Do you want to buy a gift?"*  
   - AI should have **factual personal information** to enhance user experience.

3. **User Impression ("Wow" Factor):**  
   - The AI should feel like a **loyal best friend & executive assistant**.  
   - It should infer personal details **without explicit user input**, based on **past behavior history**.

### **Question Grading Requirements**
- **Automated Grading Simplicity:**  
  - The factual data must be **verifiable** by a less-smart TEE grader.
  - Avoid **hallucinated/made-up** data by the LLM.
  - If **data is missing**, the **GOD model** should recommend a **new data connector**.

---

## **Available Data Sources**
All questions must be based **only** on the following data sources.

### **Categories & Data Sources**
1. **Social**  
   - **Twitter:** Username & basic info from OAuth.  
   - **Discord:** Username, group names, login time.  
   - **Telegram:** Username & authentication date.  
   - **Facebook/Instagram/Other:** Extracted from **Gmail receipts**.

2. **Productivity**  
   - **Gmail:** All data.  
   - **Google Calendar:** All data.

3. **Daily Life** (from **Gmail receipts** only)  
   - **Ride services:** Uber, Lyft, Waymo, Didi, Grab.  
   - **Food delivery:** Uber Eats, DoorDash, Grubhub.  
   - **Grocery delivery:** Instacart, Amazon Fresh, Costco.

4. **Shopping** (from **Gmail receipts** only)  
   - **Shopping Apps:** Amazon, Shopify, Shein, Macy's, Lululemon, etc.

5. **Web3**  
   - **Crypto Wallets:** MetaMask, Phantom, WalletConnect.  
   - **On-Chain Data:** Pulled from **3rd-party search API services**.

6. **Finance** (from **Gmail receipts** only)  
   - **Stock Brokers:** Robinhood, IB, Futu, Tiger, Charles Schwab, etc.  
   - **Crypto Exchanges:** Coinbase, Binance, Kraken, OKX, Bybit, Upbit, KuCoin, Crypto.com, etc.

7. **AI Native Chat History**  
   - **Usage history** from ChatGPT, Gemini, DeepSeek, Doubao, Perplexity, Character AI, etc.

---

## **Task Requirements**
1. **Generate 10 Questions per Data Connector Type** (for each category).  
2. **Select the Best 8 Questions per Category**  
   - Ensure questions **represent all data connectors** in the category.  
3. **Optimize for Personalization, Grading, and Predictive Power**  
   - Questions must extract **objective, verifiable, and stable** personal information.  
   - Avoid **subjective** questions (e.g., *favorite color*, which may change).  

---

## **Memory Table Structure**
Each **data connector** generates a **personal memory table**, which is:
- **Continuously updated** when new data arrives.
- **Categorized** under **Productivity/Social/Shopping/Finance/Web3/AI Native**.
- **Verified for consistency** (cross-checking between data dump & streaming APIs).
- **Formatted as Key-Value Pairs with Reference Sources**.

### **Special Tokens**
- **NE (Non-Exist):** Data does not exist after thorough verification.  
- **NA (Not Available):** Data pipeline failure (e.g., parser or LLM failure).  

> **Note:** *NE and NA do not have reference sources and do not affect scores.*
\end{lstlisting}
\end{document}